\documentclass[twocolumn]{autart}

\usepackage{graphicx}
\usepackage{caption}
\usepackage{tabularx,booktabs}
\usepackage{enumerate}
\usepackage{amssymb}
\usepackage{multicol}
\usepackage{amsmath,amssymb}
\usepackage{amsfonts}
\usepackage{bbm}
\usepackage{textcomp}
\usepackage{psfrag}
\usepackage{multimedia}
\usepackage{fancybox}

\newcommand{\E}[2]{\mathbb{E}_{#1}\left[#2\right]}

\newcommand{\Det}[1]{\mathrm{det}\left(#1\right)}
\newtheorem{Theorem}{Theorem}
\newtheorem{Proposition}{Proposition}
\newtheorem{Corollary}{Corollary}
\newtheorem{Lemma}{Lemma}
\newtheorem{Definition}{Definition}
\newtheorem{Remark}{Remark}

\usepackage[usenames,dvipsnames]{xcolor}
\definecolor{wheat}{rgb}{0.96,0.87,0.70}
\definecolor{mario}{rgb}{0.8,0.8,1}
\definecolor{seb}{rgb}{0.8,1,0.8}

\newcommand {\matr}[2]{\left[\begin{array}{#1}#2\end{array}\right]}
\newcommand {\cmatr}[2]{\left\{\begin{array}{#1}#2\end{array}\right.}
\newcommand{\tightMatr}[2]{\begin{bmatrix}#2\end{bmatrix}}

\newcommand{\vect}[1]{{\ensuremath{\boldsymbol{{#1}}}}}

\newcommand {\Tr}[1]{\mathrm{Tr}\left(#1\right)}

\newcounter{lastnote}

\begin{document} 

\begin{frontmatter}
		
	\title{
	Economic MPC of Markov Decision Processes: Dissipativity in Undiscounted Infinite-Horizon Optimal Control
	\thanksref{footnoteinfo}} 
	
	\thanks[footnoteinfo]{This paper was not presented at any IFAC 
		meeting. Corresponding author M. Zanon. This paper was partially supported by the Italian Ministry of University and Research under the PRIN'17 project ``Data-driven learning of constrained control systems" , contract no. 2017J89ARP; by ARTES 4.0 Advanced Robotics and enabling digital Technologies \& Systems 4.0, CUP: B81J18000090008; and by the Norwegian Research Council project ``Safe Reinforcement-Learning using MPC" (SARLEM).}
	
	\author[Seb]{Sebastien Gros}\ead{sebastien.gros@ntnu.no},
	\author[Mario]{Mario Zanon}
	
	\address[Seb]{Dept. of Eng. Cybernetics, Faculty of Information Technology, NTNU, Trondheim, Norway}
	\address[Mario]{IMT School for Advanced Studies Lucca, Piazza San Francesco 19, 55100, Lucca, Italy}


	\begin{abstract}
	Economic Model Predictive Control (MPC) dissipativity theory is central to discussing the stability of policies resulting from minimizing economic stage costs. In its current form, the dissipativity theory for economic MPC applies to problems based on deterministic dynamics or to very specific classes of stochastic problems, and does not readily extend to generic Markov Decision Processes. In this paper, we clarify the core reason for this difficulty, and propose a generalization of the economic MPC dissipativity theory that circumvents it. This generalization focuses on undiscounted infinite-horizon problems and is based on nonlinear stage cost functionals, allowing one to discuss the Lyapunov asymptotic stability of policies for Markov Decision Processes in terms of the probability measures underlying their stochastic dynamics. This theory is illustrated for the stochastic Linear Quadratic Regulator with Gaussian process noise, for which a storage functional can be provided explicitly. For the sake of brevity, we limit our discussion to undiscounted Markov Decision Processes.
	\end{abstract}

\end{frontmatter}

\section{Introduction}

The use of optimization-based policies is widespread in control, with the notable example of Model Predictive Control (MPC) which has gained increasing popularity in the last decades. In the deterministic setting, MPC schemes are often used to steer a system to a given feasible reference input and state. In that context, the stage cost minimized in the MPC scheme is typically convex, taking its minimum at the reference. Quadratic costs are the most common choice. This type of MPC scheme is commonly referred to as \textit{tracking MPC}.

In a control context, the system stability in closed loop with a policy is a crucial feature. In particular, an asymptotically stable policy ensures that the closed-loop system will be steered to a specific steady-state. The stability of tracking MPC schemes for deterministic systems is fairly straightforward to establish, and---under a mild controllability assumption---simply requires the MPC stage cost to be lower-bounded by a class-$\mathcal{K}_\infty$ function, with the possible addition of a terminal cost and constraint set in the finite-horizon setting. When these criteria are fulfilled, a stability argument can be easily constructed via the Lyapunov stability theory~\cite{Gruene2017,Rawlings2017}. 

Tracking MPC schemes are, however, fairly restrictive as to which stage cost can be used, and using stage costs belonging to a broader class of functions can be beneficial. Indeed, the recent literature on MPC argues for the use of \textit{economic} stage costs, representing directly the performance of the system with regard to the overall control goals, rather than the specific objective of steering the system to a given reference, see, e.g., \cite{Gros2013e,Hult2018a,Vasilj2018}. Such economic objectives often correspond to the energy, the time or the financial cost of performing a given task. It is commonly argued that a policy minimizing an economic stage cost is more conducive to maximizing the system performance than a policy optimizing a tracking stage cost can be~\cite{Rawlings2009}.

While appealing, economic stage costs typically do not satisfy the criteria required to conclude the stability of the resulting optimal control policy. To address that issue, a new stability theory has been developed, commonly referred to as \textit{dissipativity} theory in the context of economic MPC.  We ought to stress here that this dissipativity framework is very specific, as it focuses on the analysis of systems in closed-loop with optimal control policies. This framework ought not be mistaken for the general dissipativity theory for dynamical systems. The key idea behind economic MPC dissipativity theory is to transform the economic stage cost into one that is lower-bounded by a class-$\mathcal{K}_\infty$ function, while leaving the resulting policy unchanged. This transformation is often referred to as \textit{cost rotation}, and performed via a so-called \textit{storage function}. If this transformation is possible, i.e., if strict dissipativity holds, then the stability of economic MPC can be analyzed via the Lyapunov stability theory~\cite{Amrit2011a,Diehl2011,Faulwasser2018a,Grune2013a,Mueller2015,Zanon2017e}. A strong point of the economic MPC disspativity theory is that it is based on the interplay between stability and optimality, and therefore provides a natural way of discussing the stability of optimal policies without requiring the construction of the actual policy. The aforementioned dissipativity theory applies to systems having deterministic dynamics, and is not yet extended to general stochastic systems, with the notable exceptions of~\cite{Bayer2018,Sopasakis2017} which, however, only apply to rather specific settings. 

MPC for stochastic systems is often treated within the \textit{Robust MPC} (RMPC) or \textit{Stochastic MPC} (SMPC) frameworks. The former is equipped with stability theories, but the analysis is usually restricted to tracking MPC formulations, to the exception of~\cite{Bayer2016,Bayer2018}, which discuss the economic setting. However, the stability results found in the RMPC context are limited to proving the stability of the closed-loop trajectories to a set~\cite{Chisci2001,Mayne2005,Zanon2021a} under the assumption of bounded support. That set can be arbitrarily large and the behavior of the trajectories within the set is not discussed by the theory. SMPC often targets the minimization of the given stage cost in terms of its expected value taken over the stochastic predicted trajectories and the associated stability theories are less mature~\cite{Mesbah2016}.

Minimizing the expected value of a stage cost subject to stochastic trajectories is generally referred to as a Markov Decision Process (MDP). 
MDPs can be formulated in a variety of settings. 
For the sake of brevity, we will focus on discrete-time bias-optimal undiscounted MDPs over an infinite horizon and continuous state spaces, which are the closest ones to economic MPC and to the current deterministic economic MPC dissipativity setup. The extension of our results to other settings is the subject of current research.

The stability of MDPs can be arguably analyzed in the broader context of Markov Chains~\cite{Meyn2009}. Unfortunately, this framework provides results that are not easily related to optimality and therefore to the original MDPs. 
A discussion on the stability of MDPs in the context of economic MPC would therefore be beneficial, as they would bridge that gap. Unfortunately, for reasons that we detail in this paper, the direct extension of the established economic MPC dissipativity theory to MDPs is restricted to some very specific problems,  see, e.g.,~\cite{Sopasakis2017}.

As an alternative to a direct extension of the existing theory, we propose in this paper to form a generalization of the economic MPC dissipativity theory built on the measure space underlying the MDP rather than on the state space itself. This approach yields stronger results than stability to a set, and is not restricted to a specific class of stochastic dynamics. For the sake of clarity, we ought to stress here that the dissipativity theory we will develop is to be understood as a generalization of the dissipativity theory for economic MPC, rather than as a general dissipativity theory for stochastic systems investigated in, e.g.,~\cite{Rajpurohit2017,Knobloch2005}.

\textit{Contributions:} In this paper we propose a generalization of both the established dissipativity theory and the stochastic dissipativity theory of~\cite{Sopasakis2017}, and provide a stronger discussion on the stability of MDPs than convergence to a set, and a more specific stability theory than the generic discussions on the stability of Markov Chains. We show that the philosophy underlying the established dissipativity theory for economic MPC is valid in the MDP context, but its application requires generalizing the concept of stage cost and storage functions to functionals operating on the set of probability measures (or densities). The classic notion of norm must then be replaced by the notion of dissimilarity measures such as, e.g., the Kullback-Leibler divergence, the Wasserstein metric, or the total variational distance~\cite{Meyn2009}. Strong stability results follow, where dissipativity ensures that the measures underlying the MDP converge asymptotically to the steady-state optimal measure. This generalization is illustrated in the Linear Quadratic Regulator (LQR) case subject to a Gaussian process noise, for which an explicit storage functional is provided. 

The paper is organized as follows. Section \ref{sec:MDPStability} proposes a discussion of the difficulties of extending the established dissipativity theory to MDPs. Section \ref{sec:GeneralDissipativity} proposes a generalization of the classical Lyapunov-based stability arguments for optimal policies to MDPs, using a functional approach. The resulting stability theory shows that the cost function normally used in MDPs, made of the expected sum of stage costs, does not satisfy the necessary criterion for stability, and that a cost rotation is in general needed. Section \ref{sec:CostRot} generalizes the concept of rotation, dissipativity and storage function, allowing for a discussion of MDP stability in the Lyapunov context. In section \ref{sec:LQR}, these concepts are deployed in the stochastic LQR context, showing that the proposed approach is sensible. Section~\ref{sec:conclusions} concludes the paper.

\section{Markov Decision Processes and Dissipativity Theory} \label{sec:MDPStability}

In this section, we detail the Markov Decision Processes considered in this paper, and provide a brief introduction to the state of the art on dissipativity for Economic MPC, providing a framework to discuss the closed-loop stability of optimal control policies. We then discuss and detail formally why the established dissipativity theory cannot readily apply to stochastic problems, see Lemma~\ref{Lem:EquivalenceToClassic} and Remark~\ref{rem:ConsequenceOfLemma1}, and hence motivate the extension of the established dissipativity theory to MDPs, which is provided in this paper.

We consider discrete dynamical systems evolving on a continuous state space over $\mathbb R^n$, with stochastic states $\vect s_k\in\mathbb R^n$, where $k$ denotes the discrete time. The underlying measure space for $\vect s_k$ is $\mathbb R^n$ equipped with the Lebesgue measure $\upsilon$ as reference measure, and the set of Lebesgue-measurable sets as $\sigma$-algebra $\mathcal{S}$. The actions (control inputs) $\vect a_k$ are taken in the continuous space $\mathbb R^m$. We then consider stochastic dynamics defined by the conditional probability measure $\xi$:
\begin{align}
	\label{eq:markov_chain}
\xi\left[\,\mathcal B\,|\,\vect s_k,\vect a_k\right],\quad \vect s_{k+1}\in \mathcal B,
\end{align}
defining the conditional probability of observing a transition from a given state-action pair $\vect s_k,\vect a_k$ to a subsequent state $\vect s_{k+1}$ in the Lebesgue-measurable set $\mathcal B\subseteq\mathbb R^n$. 
Furthermore, we consider deterministic causal policies $\vect\pi:\,\mathbb R^n\rightarrow \mathbb R^m$ such that:
\begin{align}
	\label{eq:policy}
	\vect a_k = \vect \pi\left(\vect s_k\right).
\end{align}
We label $\mathcal M$ the set of probability measures over $\mathbb R^n$ and $\Pi$ the set of policies, i.e., $\vect\pi\in \Pi$. A policy $\vect\pi$ in closed loop with dynamics \eqref{eq:markov_chain} generates a closed-loop Markov Chain with the underlying sequence of probability measures $\rho_{k} \in \mathcal M, \ k=0,...,\infty$ describing the stochasticity of the MDP states $\vect s_k$. Hence, in the following, for each $k=0,\ldots, \infty$, $\vect s_k$ will be a sequence of random variables, and $\rho_k$ the associated sequence of probability measures, i.e., $\vect s_k \sim \rho_k$.
We then define the transition operator $\mathcal T_{\vect\pi}:\, \mathcal M \times \Pi \rightarrow \mathcal M$ as the map from a measure $\rho_k$ to its successor $\rho_{k+1}$ via \eqref{eq:markov_chain} and under policy $\vect\pi$. More specifically, $\mathcal T_{\vect\pi}$ is formally defined as \cite{Meyn2009}
\begin{align}
\label{eq:Toperator}
\rho_{k+1}(\cdot) = \mathcal T_{\vect\pi}\, \rho_{k}(\cdot) = \int \xi\left[\,\cdot\,|\,\vect s,\vect \pi\left(\vect s\right)\right]\rho_{k}\left(\mathrm d  \vect s\right) .
\end{align} 
In the following, we will restrict our policies $\vect\pi$ to be in the set $P \subseteq \Pi$ such that the integral in \eqref{eq:Toperator} is well-defined for all Borel sets. The triple $(S_{\vect{s}}, \mathcal{S}_{\vect s}, \mathbb{P}_{\vect{\pi}})$ defines the probability space associated with a Markov chain, where $S_{\vect{s}}=\prod_{k=0}^\infty \mathbb{R}^n$, with associated $\sigma$-field $\mathcal{S}_{\vect s}$, and $\mathbb{P}_{\vect{\pi}}$ is the probability measure defined by~\eqref{eq:markov_chain}-\eqref{eq:policy}~\cite{Meyn2009}.

Note that for most of the discussions in this paper, the sequence of probability measures $\rho_{0,\ldots,\infty}$ can also be interpreted as a sequence of probability densities if the measures $\rho_{0,\ldots,\infty}$ have a Radon-Nykodim derivative with respect to the Lebesgue measure.  In that case, we will assume that the associated probability densities are all in $\mathcal L^p\left(\mathbb R^n,\mathcal{S},\upsilon\right)$. The use of measures instead of densities here is a technicality aimed at providing a generic discussion. Let us label $\mathbb E_{\vect s\sim \rho_{k}}[\cdot]$ the expected value operator with respect to probability measure $\rho_{k}\in \mathcal M$. Furthermore, let us define the stage cost function $L:\,\mathbb R^n\times\mathbb R^m\rightarrow \mathbb R$. We then consider undiscounted MDPs \cite{Puterman1994} over an infinite horizon, of the classic form:
\begin{subequations}
\label{eq:MDP}
\begin{align}
J_\star\left[\rho\right] = \min_{\vect \pi\in P}&\quad \sum_{k=0}^\infty \mathbb E_{\vect s\sim \rho_{k}}\left[ L\left(\vect s,\vect \pi\left(\vect s\right)\right)- L_0\right] \label{eq:MDPCost}\\
\mathrm{s.t.} &\quad \rho_{k+1} = \mathcal T_{\vect\pi}\, \rho_{k}, \quad \rho_0 = \rho,
\label{eq:MDPDynamics}
\end{align}
\end{subequations}
where $J_\star:\, \mathcal M\rightarrow \mathbb R$, and the argument of $J_\star$, i.e., $\rho\in \mathcal M$, specifies the initial condition of the Markov Chain~\eqref{eq:MDPDynamics}. Constant $L_0\in\mathbb R$ is the optimal cost of the optimal steady-state problem
\begin{subequations}
	\label{eq:rhoinf}
	\begin{align}
		L_0 = \min_{\vect\pi\in P,\,\rho}&\quad \mathbb E_{\vect s\sim \rho}\left[ L\left(\vect s,\vect \pi\left(\vect s\right)\right)\right] \\
		\mathrm{s.t.}&\quad  \rho = \mathcal T_{\vect\pi}\, \rho, 
	\end{align}
\end{subequations}
which also delivers the corresponding  optimal steady-state measure $\rho_\star\in \mathcal M$. We assume that $\rho_\star$ exists and is unique. Problem \eqref{eq:MDP} defines an optimal policy $\vect\pi_\star:\,\mathbb R^n\rightarrow \mathbb R^m$ in $P$.

For simplicity, we assume throughout the paper that Problem~\eqref{eq:MDP} has a unique minimizer. We do not expect the theory presented hereafter to change if $\vect \pi$ solution of \eqref{eq:MDP} is not unique or defined via an infimization rather than a minimization problem. This would arguably require additional technicalities, though, which we avoid here for the sake of simplicity.

\begin{Remark}
	In order to frame Problem~\eqref{eq:MDP} in the general context of undiscounted MDPs, we observe that, in case $L_0$ is also the optimal average cost, given by 
	\begin{subequations}
		\label{eq:average_cost_MDP}
		\begin{align}
			L_0 = \min_{\vect \pi\in P}&\quad \lim_{N\to\infty}\frac{1}{N}\sum_{k=0}^{N-1} \mathbb E_{\vect s\sim \rho_{k}}\left[ L\left(\vect s,\vect \pi\left(\vect s\right)\right)\right]\\
			\mathrm{s.t.} &\quad   \rho_{k+1} = \mathcal T_{\vect\pi}\, \rho_{k}, \quad \rho_0=\rho,
		\end{align}%
	\end{subequations}%
	$\forall\,\rho$,
	then Problem~\eqref{eq:MDP} yields bias optimality, a formal definition of which is available in, e.g.,~\cite{Puterman1994}. 
	This observation is only provided to give additional context and does not impact the remainder of this paper.
\end{Remark}

We ought to observe that, while in this paper we focus on undiscounted MDPs of the form~\eqref{eq:MDP}, we expect that our framework can be readily extended to cover the discounted case by building on the ideas proposed in~\cite{Zanon2021d}. This extension is the subject of current research.

In this paper, we are interested in characterizing conditions on the MDP dynamics and stage cost $L$ such that the optimal policy $\vect\pi_\star$ solution of \eqref{eq:MDP} is stabilizing the closed-loop Markov Chain to the optimal steady-state solution of \eqref{eq:rhoinf}, i.e., such that
\begin{align}
\lim_{k\to\infty}\rho_k=\rho_\star,
\end{align}
in some sense that we will discuss. 

 We recall that in the special case where \eqref{eq:MDP} is deterministic, such that $\rho_k$ reduces to a sequence of Dirac measures, the stability of \eqref{eq:MDP} can be discussed in the framework of the established dissipativity theory for economic MPC, using the concept of storage function. In that context, a storage function $\vect\lambda:\,\mathbb R^n\rightarrow \mathbb R$ is sought, such that 
\begin{align}
\label{eq:BasicDissipativity}
L\left(\vect s_k,\vect \pi\left(\vect s_k\right)\right) - \vect\lambda\left(\vect s_{k+1}\right) + \vect\lambda\left(\vect s_k\right) \geq \varrho\left(\|\vect s_k - \vect s_\star\|\right)
\end{align} 
holds over the deterministic system trajectories for an optimal steady state $\vect s_\star$ of the system and a class-$\mathcal{K}_\infty$\footnote{We define $\mathbb{R}_{+}:=\{ \ x \in\mathbb{R} \ | \ x\geq0 \ \}$. Function $\varrho:\mathbb{R}_{+}\to\mathbb{R}_{+}$ satisfies $\varrho\in\mathcal{K}$ if it is continuous, zero at zero and strictly increasing. If additionally $\varrho\in\mathcal{K}$ is radially unbounded, then $\varrho\in\mathcal{K}_\infty$.} function $\varrho:\mathbb R_+\rightarrow \mathbb R_+$. 
Under the condition that the storage function remains bounded over the prediction horizon, the optimal value function $J_\star^\mathrm{R}$ resulting from the \textit{rotated} cost $L^\mathrm{R}:\,\mathbb R^n\times\mathbb R^m\rightarrow \mathbb R_+$ defined as
\begin{align}
\label{eq:CostRotation}
L^\mathrm{R}\left(\vect s_k,\vect \pi\left(\vect s_k\right)\right)= L\left(\vect s_k,\vect \pi\left(\vect s_k\right)\right) - \vect\lambda\left(\vect s_{k+1}\right) + \vect\lambda\left(\vect s_k\right) 
\end{align}
is a Lyapunov function for the system. The general philosophy of the dissipativity theory is to transform the stage cost $L$ of an economic optimal control problem \eqref{eq:MDP} into a new stage cost $L^\mathrm{R}$ that yields the same optimal policy, while resulting in a value function $J_\star^{\mathrm{R}}$ that is a Lyapunov function for the closed-loop trajectories. 

A direct extension of this philosophy to treat the stability of MDPs in the form \eqref{eq:MDP} is appealing. This approach has been followed in~\cite{Sopasakis2017} for Markovian switching systems, where the rotated cost is given by
\begin{align}
	\label{eq:LinearRotation}
	&\mathbb E_{\vect s\sim \rho_{k}}\left[ L^\mathrm{R}\left(\vect s,\vect \pi\left(\vect s\right)\right)\right]  = \\&\mathbb E_{\vect s\sim \rho_{k}}\left[ L\left(\vect s,\vect \pi\left(\vect s\right)\right)\right]+\mathbb E_{\vect s\sim \rho_{k},\vect s_+\sim \rho_{k+1}}\left[ \vect\lambda\left(\vect s_+\right) - \vect\lambda\left(\vect s\right) \right].\nonumber
\end{align}
and the stage cost $\mathbb E_{\vect s\sim \rho_{k}}\left[L^\mathrm{R}-L_0\right ]$ is used in \eqref{eq:MDP}. Note that if \eqref{eq:rhoinf} is formulated by replacing $L$ with $L^\mathrm{R}$, its optimal cost is still $L_0$. 
In order to prove that $J_\star^{\mathrm{R}}$ is non-increasing using the standard approach, one then needs
\begin{align}
	\label{eq:impossible_PD}
	\mathbb E_{\vect s\sim \rho_{k}}\left[ L^\mathrm{R}\left(\vect s,\vect \pi\left(\vect s\right)\right) -L_0\right ] \geq \mathbb E_{\vect s\sim \rho_{k}}\left[\varrho\left(\|\vect s - \vect s_\star\|\right)  \right ]
\end{align} 
to hold along the system trajectories.
This condition has been called \emph{strict stochastic dissipativity} in~\cite{Sopasakis2017}.
However, except for some special cases (e.g., Markovian switching systems), $L^\mathrm{R}-L_0$ cannot be non-negative everywhere, such that by construction~\eqref{eq:impossible_PD} cannot hold.
This statement is supported in a more formal way by the following lemma. Note that the lemma discusses the case of a rotated cost $L^\mathrm{R}$, but also applies to the original MDP~\eqref{eq:MDP} if one selects $\vect{\lambda}(\vect{s})=0$, such that $L^\mathrm{R}=L$. 
\begin{Lemma} \label{Lem:EquivalenceToClassic} For infinite-horizon, undiscounted MDPs on the continuous state space $\mathbb R^n$, the following statements cannot be all simultaneously true:
\begin{enumerate}[1.]
\item $L^{\mathrm{R}}\left(\vect s,\vect \pi\left(\vect s\right)\right)-L_0=0$ on a set $S_0\subset \mathbb R^n$ of zero Lebesgue measure and  $L^{\mathrm{R}}\left(\vect s,\vect \pi\left(\vect s\right)\right)-L_0 \geq  \varrho\left(\|\vect s\|_{S_0}\right)$ holds for all $\vect s\in\mathbb R^n$, where $\varrho$ is a continuous class $\mathcal{K}_\infty$ function, and $\|\cdot\|_{S_0}$ a continuous distance to $S_0$.
\item $J^R_\star[\rho_0]$ exists and is bounded for a non-empty set of initial probability measures $\rho_0$.
\item There is a $k_0\in\mathbb N_+$ and a constant $0<b<\infty$ such that for every $k\geq k_0$ the measures $\rho_k$ are equipped with probability density functions $f_k\,:\,\mathbb R^n\rightarrow \mathbb R_+$ such that $f_k(\vect s)\leq b$, $\forall\,\vect s\in \mathbb R^n$.
\end{enumerate}
\end{Lemma}
In order to keep the proof accessible, the argument is developed using calculus, making it fairly long but simple.
\begin{pf} 
By contradiction. In order for $J_\star^{\mathrm{R}}[\rho_0]$ to exist and be bounded (statement 2.), the limit
\begin{align}
\label{eq:ZeroLimit}
\lim_{k\rightarrow \infty} \mathbb E_{\vect s\sim \rho_{k}}\left[ L^\mathrm{R}\left(\vect s,\vect \pi\left(\vect s\right)\right)-L_0\right] =0
\end{align}
must hold. 
In order to prove the contradiction, let us assume that statements 1. and 3. hold.
Let us define the sub-level set $S_\alpha\subseteq\mathbb R^n$ as:
\begin{align}
	\label{eq:sub-level_set}
S_\alpha = \left\{\,\vect s\in\mathbb R^n\quad\mathrm{s.t.}\quad \varrho\left(\|\vect s\|_{S_0}\right)\leq \alpha\right\}.
\end{align}
Because $\varrho$ and $\|\cdot\|_{S_0}$ are continuous, and $\varrho\left(\|\vect s\|_{S_0}\right)= 0$ for some $\vect s$, $S_\alpha$ has a non-zero Lebesgue measure for any $\alpha > 0$. Let us further define the set $\mathcal S\subset\mathbb R^n$ as:
\begin{align}
\mathcal S = S_{\bar \alpha}\quad\text{where}\quad \bar\alpha>0\quad\text{ is s.t.}\quad \int_{S_{\bar \alpha}}\mathrm d \vect s = b^{-1},
\end{align}
whose existence is guaranteed by the continuity of $\varrho$ and  $\|\cdot\|_{S_0}$. Let us then define the probability density $f_b$ as
\begin{align}
f_b\left(\vect s\right) = \cmatr{ccc}{ b&\text{if}&\vect s\in \mathcal S\\
0&\text{if}&\vect s\notin \mathcal S}.
\end{align}
Note that $f_b$ is indeed a probability density, since by construction $\int_{\mathbb{R}^n} f_b \mathrm d \vect s = 1$. 
We will show next that density $f_b$ yields a strictly positive lower bound for $\mathbb E_{\vect s\sim \rho_{k}}\left[ L^\mathrm{R}\left(\vect s,\vect \pi\left(\vect s\right)\right)-L_0\right] $. We first observe that since $S_0$ is of zero Lebesgue measure, $\varrho\left(\|\vect s\|_{S_0}\right) > 0$ almost everywhere in $\mathcal S$, and $\mathcal S$ has a strictly positive Lebesgue measure, there is a constant $c$ such that
\begin{align}
\label{eq:Positive:fb}
\int_{\mathbb{R}^n}  \varrho\left(\|\vect s\|_{S_0}\right) f_b\left(\vect s\right)\mathrm d\vect s = b \int_{\mathcal S}  \varrho\left(\|\vect s\|_{S_0}\right) \mathrm d\vect s =  c>0
\end{align}
holds. Let us define $\Delta\left(\vect s\right)=f_k\left(\vect s\right) - f_b\left(\vect s\right)$. We observe that since $f_k$ and $f_b$ are both probability densities, equality
$\int_{\mathbb{R}^n}   \Delta\left(\vect s\right)\mathrm d\vect s = 0$
holds, such that
\begin{align}
\label{eq:Balance}
\int_{\mathcal S}  \Delta\left(\vect s\right) \mathrm d\vect s = - \int_{\mathcal S^{\mathrm c}} \Delta\left(\vect s\right)\mathrm d\vect s \leq 0,
\end{align}
where $\mathcal S^{\mathrm c}$ is the complementary set to $\mathcal S$, and where the inequality holds because for all $\vect s\in \mathcal S$
\begin{align}
\Delta\left(\vect s\right)= \underbrace{f_k\left(\vect s\right)}_{\leq b} - \underbrace{f_b\left(\vect s\right)}_{=b} \leq 0.
\end{align}
Furthermore, from the definition of $\mathcal S$, using~\eqref{eq:sub-level_set} we have that:
\begin{subequations}
\label{eq:boundL}
\begin{align}
0\leq \varrho\left(\|\vect s\|_{S_0}\right) &\leq \bar\alpha\qquad\forall\vect s\in \mathcal S, \label{eq:boundL1} \\
\varrho\left(\|\vect s\|_{S_0}\right) &> \bar\alpha\qquad\forall\vect s\in \mathcal S^\mathrm{c}. \label{eq:boundL2}
\end{align}
\end{subequations}
Using \eqref{eq:boundL} and \eqref{eq:Balance}, we then observe that 
\begin{subequations}
\label{eq:I:rhobounds}
\begin{align}
0\geq  \int_{\mathcal S}  \varrho\left(\|\vect s\|_{S_0}\right) \Delta\left(\vect s\right)\mathrm d\vect s &\geq \bar\alpha\int_{\mathcal S} \Delta\left(\vect s\right)\mathrm d\vect s, \label{eq:I:rhobounds1}\\
\int_{\mathcal S^{\mathrm c}}  \varrho\left(\|\vect s\|_{S_0}\right) \Delta\left(\vect s\right)\mathrm d\vect s &\geq \bar\alpha\int_{\mathcal S^c}\Delta\left(\vect s\right)\mathrm d\vect s  \label{eq:I:rhobounds2}\\&=  -\bar\alpha\int_{\mathcal S}\Delta\left(\vect s\right)\mathrm d\vect s \geq 0 \nonumber
\end{align}
\end{subequations}
Hence by summing \eqref{eq:I:rhobounds1}-\eqref{eq:I:rhobounds2} we observe that: 
\begin{align}
\label{eq:positiveDelta}
\int_{\mathbb{R}^n}  \varrho\left(\|\vect s\|_{S_0}\right) \Delta\left(\vect s\right)\mathrm d\vect s \geq 0.
\end{align}
Using \eqref{eq:positiveDelta}, \eqref{eq:Positive:fb} and the definition of $\Delta$ yields:
\begin{align}
\int_{\mathbb{R}^n}  \varrho\left(\|\vect s\|_{S_0}\right) f_k\left(\vect s\right) \mathrm d\vect s \geq \int_{\mathbb{R}^n}  \varrho\left(\|\vect s\|_{S_0}\right) f_b\left(\vect s\right)\mathrm d\vect s = c.
\end{align}
We can finally conclude by observing that for all $k\geq k_0$:
\begin{align}
\mathbb E_{\vect s\sim \rho_{k}}\left[ L^\mathrm{R}\left(\vect s,\vect \pi\left(\vect s\right)\right)-L_0\right] \nonumber&\geq \int_{\mathbb{R}^n} \varrho\left(\|\vect s\|_{S_0}\right) f_k\left(\vect s\right) \mathrm d\vect s \\ &\geq c > 0.
\end{align}
This last inequality is in contradiction with \eqref{eq:ZeroLimit}, such that statements 1., 2. and 3. cannot hold together. $\hfill\qed$
\end{pf}

\begin{Remark} \label{rem:ConsequenceOfLemma1}
The consequence of Lemma \ref{Lem:EquivalenceToClassic} is that the stage cost $L$ of an MDP can be rotated such that the resulting value function is a Lyapunov function in some very specific cases only, i.e.:
\begin{itemize}
\item The MDP state $\vect s_k$ converges to a set of zero Lebesgue measure in $\mathbb R^n$. If the sequence of measures $\rho_k$, $k=0,\ldots,\infty$ admits probability densities, these densities are unbounded as $k\rightarrow\infty$. If, e.g., $S_0$ is a point in $\mathbb R^n$, the limit of the sequence $\rho_k$, $k=0,\ldots,\infty$ is a Dirac measure. This requires the MDP dynamics to have very specific properties, such as, e.g., vanishing perturbations. Another instance in which this situation can occur is if~\eqref{eq:markov_chain} represents a Markovian switching system, see~\cite{Sopasakis2017}. 
\item The cost rotation makes $L^\mathrm{R}\left(\vect s,\vect\pi_\star(\vect s)\right)-L_0=0$ on a measurable set $S_0 \subseteq \mathbb R^n$. The stability discussion is then limited to stating that the sequence of probability measures $\rho_k$, with $k=0,\ldots,\infty$ will converge to a probability measure $\rho_\infty$ that has its entire support in $S_0$, i.e., $\rho_\infty(S_0)=1$. Concretely, the MDP state $\vect s_k$ will then converge to $S_0$ with probability 1, but nothing can be said of the evolution of  $\vect s_k$ within $S_0$. Such situations are typical of robust MPC, as studied in, e.g.,~\cite{Bayer2016,Bayer2018}.
\end{itemize}
A direct extension of the established dissipativity theory cannot apply to a more general context. 
\end{Remark}
In this paper, we will show that the issues detailed above do not stem from the philosophy underlying the established dissipativity theory, but simply from considering a cost rotation \eqref{eq:LinearRotation} that is restricted to being linear in the probability measures $\rho_k$.
\section{Lyapunov Stability for MDPs} \label{sec:GeneralDissipativity}
The previous section details why the established dissipativity theory does not readily apply to stochastic problems. In this section, we will show that the issue can be addressed within the general philosophy of the established dissipativity theory, but that it requires an extension where the classic state-space of deterministic systems must be replaced by the set of probability measures underlying the stochastic systems. In order to develop this extension, let us generalize MDP \eqref{eq:MDP} into the following optimal control problem:
\begin{subequations}
\label{eq:FOCP}
\begin{align}
V_\star\left[ \rho\right] = \min_{\vect \pi\in P}&\quad \sum_{k=0}^\infty \mathcal L\left[ \rho_k,\vect \pi\right] \\
\mathrm{s.t.} &\quad  \rho_{k+1} = \mathcal T_{\vect\pi}\, \rho_{k}, \quad \rho_0 = \rho, 
\label{eq:MDPDynamics2}
\end{align}%
\end{subequations}%
where $V_\star:\, \mathcal M\rightarrow \mathbb R$, and the argument of $V_\star$, i.e., $\rho$, specifies the initial condition of the Markov Chain \eqref{eq:MDPDynamics2}. The stage cost $\mathcal L\in \mathcal M \times P\rightarrow \mathbb R$ is a (possibly nonlinear) functional over the probability measures $ \rho_k$,
 and the policy $\vect \pi$. 
We assume here that $V_\star\left[\rho \right] $ is finite for a non-empty set of measures $\rho$. One can readily observe that Problem \eqref{eq:MDP} is a special case of \eqref{eq:FOCP}, obtained by selecting
\begin{align}
\label{eq:MDP:FOCP}
\mathcal L\left[ \rho_k,\vect \pi\right] = \mathbb E_{\vect s\sim \rho_{k}}\left[ L\left(\vect s,\vect \pi\left(\vect s\right)\right)-L_0\right] .
\end{align}%
Note that the condition $\mathcal L\left[ \rho_\star,\vect \pi_\star\right] = 0$ is required for $V_\star$ to be bounded. In the specific case of Problem~\eqref{eq:MDP}, this can further be seen by using the definition of Equation~\eqref{eq:rhoinf} in~\eqref{eq:MDP:FOCP}.
We will see in this paper that the freedom of using more general functionals than \eqref{eq:MDP:FOCP} for $\mathcal L$ is the key to generalizing the economic NMPC dissipativity theory to MDPs.

We detail next how \eqref{eq:FOCP} makes it possible to build a fairly straightforward and classic Lyapunov stability result on the set of probability measures. To that end, let us introduce the following key concepts.
\begin{Definition}[Dissimilarity measure]
Let us define a dissimilarity measure $D\left(\cdot\,||\cdot\right): {\mathcal{R}}\times {\mathcal{R}}\to\mathbb{R}$ as an application from a subset of probability measures to the real positive numbers, such that:
\begin{align}
D\left(\rho\,||\,\rho^\prime\right) \geq 0\quad\text{and}\quad D\left(\rho\,||\,\rho\right) = 0,\quad\forall\,\rho,\rho^\prime\in {\mathcal{R}}\, ,
\end{align}%
where ${\mathcal{R}}\subseteq \mathcal M$ is (a subset of) the set of probability measures. 
\end{Definition}
A useful example of dissimilarity measures is the Kullback-Leibler divergence ($D_\mathrm{KL}$) defined as
\begin{align}
D_\mathrm{KL}\left(\rho\,||\,\rho^\prime\right) = \int \log \frac{\mathrm d \rho}{\mathrm d\rho^\prime}\mathrm d\rho= \int f(\vect s)\log \frac{f(\vect s)}{f^\prime(\vect s)}\mathrm d\vect s ,
\end{align}%
where $\frac{\mathrm d \rho}{\mathrm d\rho^\prime}$ is the Radon-Nykodim derivative of $\rho$ with respect to $\rho^\prime$ and the second equality holds if $\rho,\,\rho^\prime$ have underlying probability densities $f,\,f^\prime$. Other useful examples in control are the Wasserstein metric, and the total variational distance~\cite{Meyn2009}. The notion of stability on the set of probability measures can then be formalized as follows.
\begin{Definition}[$D$-Stability]
	A Markov Chain is $D$-stable with respect to probability measure $\rho_\star$ and dissimilarity measure $D$ if, for any $\epsilon>0$ there exists a $\delta(\epsilon)>0$ such that $D\left(\rho_0\,||\,\rho_\star\right) < \delta$ implies $D\left(\rho_k\,||\,\rho_\star\right) < \epsilon$ for all $k\geq0$. If, moreover, the probability measure $\rho_\star$ is $D$-attractive, i.e.,
	\begin{align}
		\lim_{k\rightarrow \infty}D\left(\rho_k\,||\,\rho_\star\right) = 0, \label{eq:ConvergenceInDissimilarity}
	\end{align}%
	holds, then the Markov Chain is $D$-asymptotically stable.
	
	$\hfill\square$
\end{Definition}
Note that $D$-stability is introduced as a practical way to encompass many commonly used stability concepts. If, e.g., $D$ is the total variational distance, then the obtained stability is often referred to as ergodicity~\cite{Meyn2009}. If, e.g., $D$ is the expected value under $\rho$ of the square of $\vect{s}$, then one obtains asymptotic stability for deterministic systems~\cite{Rawlings2017} and mean square stability for stochastic systems~\cite{Sopasakis2017}.

The next theorem formalizes the stability of OCP \eqref{eq:FOCP} on the set of probability measures, following the same arguments as classic Lyapunov stability for optimal policies over deterministic problems.
\begin{Theorem}\label{thm:asymptotic_stability} Assume that the inequalities
\begin{subequations}
\begin{align}
\mathcal L\left[\rho_k,\vect \pi_\star\right]  &\geq \varrho_1 \left( D\left(\rho_k\, ||\, \rho_\star\right) \right), \label{eq:Lbound} \\
V_\star[ \rho_k] &\leq \varrho_2\left(D\left( \rho_k\,||\,\rho_\star\right)\right), \label{eq:Vbound}
\end{align}%
\end{subequations}%
hold for some class-$\mathcal{K}_\infty$ functions $\varrho_{1,2}$ and for all $\rho\in\Xi\subseteq {\mathcal{R}}$, where set $\Xi$ is a non-empty set such that $V_\star< \infty$ on $\Xi$.
Then the Markov chain is $D$-asymptotically stable with respect to the probability measure $\rho_\star$ and dissimilarity measure $D$ for any $\rho_0\in\Xi$. \flushright$\square$
\end{Theorem}
\begin{Remark} 
	\label{rem:controllability}
	Note that assumption \eqref{eq:Vbound} corresponds to a standard assumption in the context of MPC---referred to as a form of weak controllability~\cite{Rawlings2017}---which requires that the value function is upper-bounded by a class-$\mathcal{K}_\infty$ function of a norm of $\vect s-\vect s_\star$.
\end{Remark}
\begin{Remark} Condition \eqref{eq:Lbound} can be mistaken for the simple requirement that the cost functional $\mathcal L$ should correspond to a stochastic tracking MPC scheme. This interpretation is, however, not necessarily correct. Indeed, a stochastic tracking MPC scheme would typically use a cost functional of the form
\begin{align}%
\mathcal L\left[ \rho_k,\vect \pi_\star\right] = \frac{1}{2}\mathbb E_{\vect s\sim\rho_k}\left[\vect s^\top Q\vect s + \vect \pi_\star\left(\vect s\right)^\top R \vect \pi_\star\left(\vect s\right) \right],
\end{align}
which typically does not satisfy condition \eqref{eq:Lbound} unless $\rho_\star$ is a Dirac measure centered at $\vect s=0$. The latter requires that some very specific properties are satisfied by the system dynamics \eqref{eq:markov_chain}, and hence does not hold in general. See Lemma \ref{Lem:EquivalenceToClassic} and the following remarks.
\end{Remark}
\begin{pf} (of Theorem \ref{thm:asymptotic_stability})
We first observe that because $V_\star$ is bounded on $\Xi$, $\Xi$ is positive invariant  for system $\xi$ defined in~\eqref{eq:markov_chain} under policy $\vect{\pi}_\star$ solving~\eqref{eq:FOCP} and
\begin{align}
V_\star[\rho_{k+1}] - V_\star[\rho_k] = -\mathcal L\left[ \rho_k,\vect \pi_\star\right]  \leq -\varrho_1\left(D\left(\rho_k\,||\,\rho_\star\right)\right) \label{eq:Vdecrease}
\end{align}%
holds on $\Xi$. Furthermore, we observe that from \eqref{eq:Lbound}, the bound
\begin{align}
V_\star[\rho_k] \geq \mathcal L\left[ \rho_k,\vect \pi_\star\right] \geq \varrho_1 \left( D\left(\rho_k\, ||\, \rho_\star\right) \right) \geq 0 \label{eq:Vlowerbound}
\end{align}%
holds for any $\rho_k\in\Xi$. Hence $V_\star[\rho_k] \geq 0$ is bounded and monotonically decreasing on $\Xi$, such that it must converge to a finite positive value $\bar V$ as $k\rightarrow \infty$. We then need to prove that $\bar V = 0$. To that end, consider $\delta,\epsilon>0$ selected as
\begin{align}
D\left ( \rho_0 \,||\, \rho_\star \right )\leq \delta,\qquad \epsilon = \varrho_1^{-1}(\varrho_2(\delta)),
\end{align}%
such that 
\begin{align}
V_\star[\rho_0]\leq \varrho_2(\delta)=\varrho_1(\epsilon).
\end{align}%
Using \eqref{eq:Vdecrease} and \eqref{eq:Vlowerbound}, we observe that for all $k$:
\begin{align}
D\left(\rho_k\, ||\,  \rho_\star \right) &\leq \varrho_1^{-1}(V_\star[\rho_k]) \leq \varrho_1^{-1}(V_\star[\rho_0]) \nonumber \\&\leq \varrho_1^{-1}( \varrho_1(\epsilon) )=\epsilon, \label{eq:D:upperbound}
\end{align}%
which proves stability. In order to prove attractivity, we proceed by contradiction. 
Assume that
\begin{align}
\lim_{k\rightarrow\infty} V_\star[\rho_k] = \bar V>0,
\end{align}%
then using \eqref{eq:Vbound} and \eqref{eq:D:upperbound}, the inequalities
\begin{align}
\varrho^{-1}_2(\bar V) \leq \lim_{k\rightarrow\infty} D\left(\rho_k\, ||\, \rho_\star \right)  \leq \varrho^{-1}_1(\bar V)
\end{align}%
hold. 
Using \eqref{eq:Vdecrease} we obtain:
	\begin{align}
		V_\star[\rho_k] \leq V_\star[\rho_0] - \sum_{j=0}^k \varrho_1\left (D\left(\rho_j\, ||\, \rho_\star \right) \right ).
	\end{align}%
	Since $D\left(\rho_k\, ||\, \rho_\star \right) $ converges to the interval $[\varrho^{-1}_2(\bar V),\varrho^{-1}_1(\bar V)]$, then
\begin{align}
\lim_{k\rightarrow \infty} \varrho_1\left (D\left(\rho_k\, ||\, \rho_\star \right) \right ) \geq \varrho_1\left (\varrho^{-1}_2(\bar V)\right ) > 0,
\end{align}%
such that $V_\star[\rho_k] \to -\infty$ as $k\to \infty$, which is in contradiction with \eqref{eq:Vlowerbound}. Consequently, $\bar V=0$, and~\eqref{eq:ConvergenceInDissimilarity} 
must hold.$\hfill\qed$
\end{pf}

Note that the stability result of this theorem can carry several meanings, depending on the properties of $D$. If, e.g., $D(\rho_1||\rho_2)= \Big\| \E{\vect{s}\sim\rho_1}{\vect{s}}-\E{\vect{s}\sim\rho_2}{\vect{s}} \Big\|$, then only the expected value of the state is guaranteed to converge. In case the selected dissimilarity measure carries stronger properties, stronger stability results arise. Let us detail a useful special case in the next corollary.
\begin{Corollary}
	\label{cor:as_stab}
	Assume that the assumptions of Theorem~\ref{thm:asymptotic_stability} hold, and the dissimilarity measure $D\left(\rho\,||\,\rho_\star\right)$ is such that $D\left(\rho\,||\,\rho_\star\right) = 0$ implies that $\rho=\rho_\star$ almost everywhere. Then
	\begin{align}
		\lim_{k\rightarrow \infty} \rho_k\left(\cdot\right) = \rho_\star\left(\cdot\right)  \label{eq:ConvergenceInDensity}
	\end{align}%
holds almost everywhere.\flushright$\square$
\end{Corollary}
\begin{pf}
	The limit~\eqref{eq:ConvergenceInDissimilarity} follows from Theorem~\ref{thm:asymptotic_stability}. By the properties assumed on the dissimilarity measure, this directly entails \eqref{eq:ConvergenceInDensity}. $\hfill\qed$
\end{pf}
Examples of dissimilarity measure satisfying Corollary~\ref{cor:as_stab} include $D_\mathrm{KL}$ and the total variational distance. It may be useful here to discuss what form of stability is established in Theorem~\ref{thm:asymptotic_stability}. Stability proofs in the context of Classic MPC and Economic MPC discuss the behavior of single trajectories, starting from arbitrary initial conditions in a set, and proves the convergence to an optimal steady-state. In Robust MPC one discusses the behavior of all possible stochastic trajectories, and proves the convergence to a set, without describing the behavior inside that set. In contrast, Theorem~\ref{thm:asymptotic_stability} discusses the behavior of trajectories by showing that their asymptotic behavior is to be distributed according to a distribution with zero dissimilarity with respect to the optimal steady-state measure of the MDP. For suitably selected dissimilarity measures (see, e.g., Corollary~\ref{cor:as_stab}), this entails that these two distributions must coincide almost everywhere.

We now turn to discussing how the stability of the MDP resulting from a generic stage cost functional $\mathcal L$ can be discussed in terms of \eqref{eq:Lbound} via functional cost rotations.

\section{Functional Cost Rotations } \label{sec:CostRot}
\label{eq:CostModif}
Making a Lyapunov stability argument on problem \eqref{eq:FOCP} requires the cost functional $\mathcal L\left[ \rho_k,\vect \pi\right]$ to satisfy \eqref{eq:Lbound}. Following the arguments of Lemma \ref{Lem:EquivalenceToClassic}, one can readily observe that, in general, for MDP \eqref{eq:MDP} to be well-posed, the MDP stage cost {$L-L_0$} cannot be strictly positive. As a result, when recasting a given MDP \eqref{eq:MDP} in its equivalent functional form \eqref{eq:FOCP} using identity \eqref{eq:MDP:FOCP}, the resulting functional stage cost $\mathcal L\left[ \rho,\vect \pi\right]$ cannot be positive for all probability measures $\rho$, such that \eqref{eq:Lbound} cannot hold. This challenges by construction the extension of classical Lyapunov stability to general MDPs. As a result, a classic rotation in the form \eqref{eq:LinearRotation} is not applicable in general.

Fortunately, it is possible to tackle these difficulties by adopting a more general cost rotation than \eqref{eq:LinearRotation}. More specifically, we will consider functional cost rotations of the form:
\begin{align}
\label{eq:CostModification}
 {\mathcal L}^\mathrm{R}\left[ \rho_k,\vect \pi\right] = \mathcal L\left[ \rho_k,\vect \pi\right] - \lambda\left[\rho_{k+1}\right] + \lambda\left[\rho_k\right],
\end{align}%
where $\lambda:\, \mathcal M\rightarrow \mathbb R$ is a (possibly) nonlinear functional. Rotation \eqref{eq:LinearRotation} is then a special case of \eqref{eq:CostModification}, where the form
\begin{align}
\label{eq:LinearLambda}
\lambda\left[\rho_{k}\right] = \mathbb E_{\vect s\sim \rho_{k}}\left[ \lambda\left(\vect s\right)\right] 
\end{align}%
is imposed. Following the arguments presented above, the form \eqref{eq:LinearLambda} is in general not able to deliver a Lyapunov function.

Similarly to classical cost rotations, we observe that \eqref{eq:CostModification} leaves the policy solution of \eqref{eq:FOCP} unchanged, as long as $\lambda[\rho_k]$ is bounded for all $k$. A generalized dissipativity criterion can then be formulated as follows.
\begin{Definition}[Functional Strict Dissipativity]\label{def:FSD}
	There exists a functional $ \lambda:\, \mathcal M\rightarrow \mathbb R$ and a class-$\mathcal{K}_\infty$ function $\varrho$ such that ${\mathcal L}^\mathrm{R}\left[ \rho_k,\vect \pi\right]$ defined by \eqref{eq:CostModification} satisfies \eqref{eq:Lbound}, i.e.,
	\begin{align}
		\label{eq:Dissipativity}
\mathcal L\left[\rho_k,\vect\pi\right]- \lambda\left[\rho_{k+1}\right] +  \lambda\left[\rho_k\right] &\geq \varrho \left( D\left(\rho_k\, ||\, \rho_\star\right) \right) 
	\end{align}%
	holds for all $\rho_k\in{\mathcal R}$ such that $V_\star\left(\rho_k\right)$ is finite.
\end{Definition}
As we will prove next, the functional dissipativity criterion \eqref{eq:Dissipativity} then yields $D$-asymptotic stability. Indeed, let us define a rotated problem as:
\begin{subequations}
	\label{eq:rotatetd_FOCP}
	\begin{align}
		V_\star^\mathrm{R}\left[\rho\right] = \min_{\vect \pi \in P}& \quad \lim_{N\to\infty} \sum_{k=0}^{N-1} \mathcal L^\mathrm{R}\left[ \rho_k,\vect \pi\right] + \lambda\left[\rho_{N}\right] \label{eq:CostRotatedMDP}\\
		\mathrm{s.t.} &\quad \rho_{k+1} = \mathcal T_{\vect\pi}\, \rho_{k}, \quad \rho_0 = \rho, 
		\label{eq:MDPDynamics3}
	\end{align}%
\end{subequations}%
where $V_\star^\mathrm{R}:\, \mathcal M\rightarrow \mathbb R$, and the argument of $V^\mathbb{R}_\star$, i.e., $\rho$, specifies the initial condition of the Markov Chain \eqref{eq:MDPDynamics3}. We then establish $D$-asymptotic stability in the next theorem. Under the assumption that functional strict dissipativity holds for a bounded storage functional $\lambda$, the existence of the limit in \eqref{eq:CostRotatedMDP} will become clear in equation~\eqref{eq:limit_existence} in the proof of the next theorem.
\begin{Theorem} \label{Th:Stability}
	Assume that there exists a storage functional $\lambda$ bounded from above and below, and satisfying~\eqref{eq:Dissipativity}. Assume moreover that
	\begin{align}
		\label{eq:controllability}
		V_\star^\mathrm{R}[\rho_k] \leq \varrho_2(D\left(\rho_k\,||\,\rho_\star\right)).
	\end{align}
	with $V_\star^\mathrm{R}:\,\mathcal M\rightarrow \mathbb R$ defined in \eqref{eq:rotatetd_FOCP}. Then, the rotated problem~\eqref{eq:rotatetd_FOCP} and the original problem~\eqref{eq:FOCP} deliver the same optimal policy. Moreover, the Markov chain is $D$-asymptotically stable with respect to the probability measure $\rho_\star$ and dissimilarity measure $D$.
\end{Theorem}
\begin{Remark}
	We observe that~\eqref{eq:controllability} can be interpreted as a controllability assumption, since it holds whenever $\rho_k$ can be steered to $\rho_\star$ (in the sense of the dissimilarity measure $D$) with finite cost. This is equivalent, mutatis mutandis, to the deterministic case, see Remark~\ref{rem:controllability}.
\end{Remark}
\begin{Remark}
	A reader well acquainted to the dissipativity theory for economic MPC will recognize in Theorem \ref{Th:Stability} a generalization of the theory applicable in the deterministic case, where a criterion similar to \eqref{eq:Dissipativity} and a bound similar to \eqref{eq:controllability} entail the convergence of the system state to the optimal steady-state. Theorem \ref{Th:Stability} extends these concepts to the convergence of the system in the sense of the probability measures underlying the dynamics rather than in the sense of the states themselves. 
\end{Remark}

\begin{pf} (of Theorem \ref{Th:Stability})
	The first claim follows from standard arguments, since boundedness of $\lambda[\rho_k]$ entails
	\begin{align}
		\label{eq:limit_existence}
		&\sum_{k=0}^{N-1} \mathcal L^\mathrm{R}\left[ \rho_k,\vect \pi\right] + \lambda\left[\rho_{N}\right] = \sum_{k=0}^{N-1} \mathcal L\left[ \rho_k,\vect \pi\right] + \lambda\left[\rho_{0}\right].
	\end{align}
	By taking the limit $N\to\infty$, (which exists if the original problem~\eqref{eq:FOCP} is well-posed, since $\lambda$ is bounded by assumption) the cost of the rotated and original problem only differ by the constant $\lambda\left[\rho_{0}\right]$, for any evolution of the density $\rho_k$ which satisfies the Markov chain~\eqref{eq:markov_chain}. Consequently, we obtain $V_\star^\mathrm{R}\left[\rho_0 \right] = V_\star\left[\rho_0 \right] + \lambda\left[\rho_{0}\right]$.
	
	We now observe that $\mathcal{L}^\mathrm{R}$ and $V_\star^\mathrm{R}$ satisfy the assumptions of Theorem~\ref{thm:asymptotic_stability}. Consequently, the rotated problem yields a policy guaranteeing that the closed-loop system satisfies $D$-asymptotic stability with respect to density $\rho_\star$. Because the optimal policies of the rotated and original problem coincide, this proves the second claim.
	 $\hfill\qed$
\end{pf}
The existence of a bounded functional $\lambda$ satisfying \eqref{eq:Dissipativity} entails the stability of \eqref{eq:FOCP} in the set of probability measures.  We will show next that such a storage function exists in the LQR case, and can be explicitly provided, hence giving credence to this concept. In line with Lemma \ref{Lem:EquivalenceToClassic}, the resulting modified cost functional does not take the linear form \eqref{eq:LinearLambda}. It is not trivial, and its derivation is fairly technical.

\section{The LQR Case} \label{sec:LQR}
In this section, we will develop a storage functional $\lambda[\rho]$ satisfying \eqref{eq:Dissipativity} for the LQR case with Gaussian process noise and for the Kullback-Leibler divergence $D_\mathrm{KL}$. Most proofs are provided in the Appendix. We consider the dynamics:
\begin{align}
	\label{eq:linear_mdp}
	\vect{s}_+ = A\vect{s} + B\vect{a} + \vect{w},
\end{align}
where $\vect s\in\mathbb R^n$ and $\vect{w}\sim \mathcal{N}(\vect{0},\Sigma_{\vect w})$  i.i.d., $\E{}{\vect{w}\vect{s}^\top}=0$, $\E{}{\vect{w}\vect{a}^\top}=0$ and we consider the stage cost
\begin{align}
\label{LQR:cost}
	L(\vect{s},\vect{a}) = \tightMatr{c}{\vect{s} \\ \vect{a}}^\top H \tightMatr{c}{\vect{s} \\ \vect{a}}, \quad H=\tightMatr{ll}{T & U^\top \\ U & R} \succ0,
\end{align}
For $\rho_0\sim\mathcal N\left(\vect\mu_0,\Sigma_0\right)$, the dynamics of the system are given by $\rho_k\sim\mathcal N\left(\vect\mu_k,\Sigma_k\right)$ where the mean and covariance dynamics read as:
\begin{subequations}
 \label{eq:LQR:Dynamics}
\begin{align}
\vect\mu_{k+1} &= A_\mathrm{c}\vect\mu_k, \\
\Sigma_{k+1} &= A_\mathrm{c}\Sigma_k A_\mathrm{c}^\top + \Sigma_{\vect w}, \label{eq:Sigma:Dynamics}
\end{align}
\end{subequations}
and where $A_\mathrm{c} = A-BK$, and $K$ is the regular LQR matrix gain associated to $A,B,H$. Furthermore, we have
\begin{align}
 \mathbb E_{\vect s\sim\rho_k}\left[ L\left( \vect s,\vect \pi\left(\vect s\right)\right) \right] =& \mathrm{Tr}\left(W\Sigma_k\right) + \vect\mu_k^\top W\vect\mu_k ,\\
  D_{\mathrm{KL}}\left(\rho\, ||\, \rho_\star\right) =&\frac{1}{2}\left(\mathrm{Tr}\left(\Sigma_{\infty}^{-1}\Sigma_k\right) + \vect\mu_k^\top \Sigma_{\infty}^{-1}\vect\mu_k - n \right.\nonumber\\
  &\left.+ \log \mathrm{det}\left(\Sigma_{\infty}\right) -\log\mathrm{det}\left(\Sigma_k\right)\right),\nonumber
  \end{align}
where $n$ is the dimension of $\vect s$,
\begin{align}
W=\matr{c}{I \\ -K}^\top H\matr{c}{I \\ -K},
\end{align}
and $\rho_\star$ has the mean $\vect\mu_\infty = 0$ and its variance is the solution of the Lyapunov equation:
\begin{align}
	\label{eq:ss_covariance}
\Sigma_{\infty} &= A_\mathrm{c}\Sigma_\infty A_\mathrm{c}^\top + \Sigma_{\vect w}.
\end{align}
Note that~\eqref{eq:ss_covariance} has a solution only if $A_\mathrm{c}$ is stabilizing, which also entails that $\vect{\mu}_\infty=0$. 
We observe that for the MDP to be well posed
\begin{align}
L_0 = \mathrm{Tr}\left(W\Sigma_\infty\right)
\end{align}
must hold. We then observe that $\mathcal L$, as defined by \eqref{eq:MDP:FOCP} reads as
\begin{align}
\label{eq:L:LQR}
  \mathcal L\left[\rho_k,\vect\pi\right] &= \vect\mu_k^\top W\vect\mu_k + \mathrm{Tr}\left(W\left(\Sigma_k-\Sigma_\infty\right)\right).
\end{align}

One can observe that \eqref{eq:L:LQR} does not necessarily satisfy \eqref{eq:Lbound}, such that a rotation of the LQR stage cost, as per Section \ref{sec:CostRot}, is required in order for the optimal cost $V_\star$ associated to the LQR problem, defined according to \eqref{eq:FOCP}, to be a Lyapunov functional. Before delivering the storage functional associated to the LQR problem, the following section establishes some basic results confirming the convergence of the LQR problem under $ D_{\mathrm{KL}}$ independently of the proposed theory. 

\subsection{Convergence under $ D_{\mathrm{KL}}$ }\label{sec:DKL:convergence}
The stability of the stochastic LQR in the $D_{\mathrm{KL}}$ sense can be established by the theory proposed above, via providing a storage functional $\lambda$, and as a result a Lyapunov functional for the problem. This is the approach used by the established dissipativity theory in economic MPC, allowing one to build a Lyapunov function directly from the cost and dynamics of the problem at hand. We will present this approach in Section \ref{sec:StorageLQR}. Unlike general optimal control and MPC problems, the LQR problem admits a simple and explicit policy, allowing one to explicitly describe the closed-loop dynamics and discuss their stability directly. In this section, we will adopt this approach first and confirm that $D_\mathrm{KL}\left(\rho_k||\rho_\star\right) $ is monotonically decreasing under the LQR closed-loop dynamics. This will require several technical results that will also be needed for building a storage functional. The proofs of Lemma \ref{Lemma:Bazdmeg}, Lemma \ref{Lemma:ConvergenceInEigenvalues}, Proposition \ref{Prop:Generic:Dissimilarity}, Lemma \ref{Lemma:someMoreShit}, and Theorem \ref{LQR:Stability} are provided in the Appendix. In order to proceed, let us introduce first a useful technical lemma.
\begin{Lemma} \label{Lemma:Bazdmeg} Consider a full-rank matrix $M\in\mathbb R^{n\times n}$ such that its maximum singular value, i.e., $\sigma_{\max}\left(M\right)$ is less than $1$, and consider a symmetric (possibly indefinite) matrix $\Delta\in\mathbb R^{n\times n}$. Consider $\Lambda_{1,\ldots,n}(\cdot)$ the ordered eigenvalues of an $\mathbb R^{n\times n}$ matrix $\cdot$, where the indexing denotes that order. Then the following holds: 
\begin{align}
\label{eq:BazdmegInequalities}
\Lambda_i\left( M\Delta M^\top \right) = \alpha_i\Lambda_i\left(\Delta \right),\quad i=1,\ldots,n
\end{align} 
for a sequence $\alpha_{i}>0$, $i = 1,\ldots,n$ with $\alpha_i \leq \sigma_{\max}\left(M\right)$. \flushright$\square$
\end{Lemma} 
This lemma will be instrumental in showing the convergence of the LQR problem in the $D_{\mathrm{KL}}$ sense, i.e.:
\begin{align}
D_{\mathrm{KL}}\left(\rho_{k+1}\, ||\, \rho_\star\right) < D_{\mathrm{KL}}\left(\rho_k\, ||\, \rho_\star\right).
\end{align}
In order to obtain this result, the following lemma will be useful, and follows fairly directly from Lemma~\ref{Lemma:Bazdmeg}.
\begin{Lemma} \label{Lemma:ConvergenceInEigenvalues} Under dynamics \eqref{eq:Sigma:Dynamics}, the ordered eigenvalues of $\Sigma_\infty^{-1}\Sigma_k$, i.e., $\Lambda_{1,\ldots,n}\left(\Sigma_\infty^{-1}\Sigma_k\right)$
converge monotonically to $1$ without changing sign. More specifically:
\begin{align}
\label{eq:Contraction:Eigen}
\Lambda_i\left( \Sigma_\infty^{-1}\Sigma_{k+1} \right) - 1 = \alpha_i\left(\Lambda_i\left(\Sigma_\infty^{-1}\Sigma_{k} \right) -1\right),
\end{align}
for a sequence $\alpha_{1,\ldots,n}\in\mathbb R_+$ with 
\begin{align}
\alpha_i \leq \sigma_{\max}\left(\Sigma_{\infty} ^{-\frac{1}{2}}A_\mathrm{c} \Sigma_{\infty} ^{\frac{1}{2}}\right)<1,\qquad\,\forall\,i.
\end{align}
 \flushright$\square$
\end{Lemma}
Using Lemma \ref{Lemma:ConvergenceInEigenvalues}, the monotonic decreasing of a class of dissimilarity measures under the dynamics \eqref{eq:LQR:Dynamics} is {established} next.
\begin{Proposition} \label{Prop:Generic:Dissimilarity}
Consider any dissimilarity measure $D\left(\rho_k\,||\,\rho_\star\right)$ that can be expressed in the form:
\begin{align}
\label{eq:Generic:Dissimilarity}
D\left(\rho_k\,||\,\rho_\star\right) = c + \vect\mu_k^\top \Sigma_\infty^{-1}\vect\mu_k + \sum_{i=1}^n \zeta\left(\Lambda_i\left(\Sigma_\infty^{-1}\Sigma_k\right)\right),
\end{align}
for some function $\zeta:\,\mathbb R\rightarrow \mathbb R_+$ that is strictly increasing away from $1$, and some constant $c$. Then $D\left(\rho_k\,||\,\rho_\star\right)$ is strictly decreasing under dynamics \eqref{eq:LQR:Dynamics}. \flushright$\square$
\end{Proposition}
Proposition \ref{Prop:Generic:Dissimilarity} establishes that in the stochastic LQR case, the measures underlying the stochasticity of the state space converge in the sense of an entire class of dissimilarity measures (including $D_\mathrm{KL}$). This is a direct result not relying on the proposed functional dissipativity theory. However, the mathematical argument establishing Proposition \ref{Prop:Generic:Dissimilarity} is central in developing a storage function showing the functional dissipativity of LQR.

Proposition \ref{Prop:Generic:Dissimilarity} applies to several dissimilarity measures including $D_{\mathrm{KL}}$ and the Wasserstein metric. For the sake of simplicity, we focus on $D_{\mathrm{KL}}$ in the following.
\begin{Corollary} 
	Dynamics \eqref{eq:LQR:Dynamics} converge monotonically under $D_{\mathrm{KL}}$, i.e.,
\begin{align}
D_{\mathrm{KL}}\left(\rho_{k+1}\,||\, \rho_\star\right) \leq D_{\mathrm{KL}}\left(\rho_k\,||\, \rho_\star\right), 
\end{align}
and the inequalities are strict for $\rho_k \neq \rho_\star$. \flushright$\square$
\end{Corollary} 
\begin{pf} 
	We observe that 
\begin{align}
 D_{\mathrm{KL}}\left(\rho_k\, ||\, \rho_\star\right) &=\frac{1}{2}\vect\mu_k^\top \Sigma_{\infty}^{-1}\vect\mu_k     \\&+\frac{1}{2}\sum_{i=1}^n \left(\Lambda_i\left(\Sigma_{\infty}^{-1}\Sigma_k\right)  -\log\Lambda_i\left(\Sigma_{\infty}^{-1}\Sigma_k\right)-1\right), \nonumber
\end{align} 
and we observe that the scalar function 
\begin{align}
\zeta(x) = x - \log x - 1
\end{align}
is monotonically increasing away from $1$. Since $ D_{\mathrm{KL}}$ differs from the form \eqref{eq:Generic:Dissimilarity} only by a factor $\frac{1}{2}$, the monotonic decrease is conserved.$\hfill\qed$
\end{pf}

\subsection{A Storage Functional for $D_\mathrm{KL}$} \label{sec:StorageLQR}
Section \ref{sec:DKL:convergence} shows the stability of the stochastic LQR closed-loop trajectories under $D_\mathrm{KL}$ directly. This can be done in the LQR case where the closed-loop dynamics are known explicitly. For general optimal control and MPC problems, the optimal policy is typically not explicitly known and the same approach cannot be used. Dissipativity theory then allows one to study the stability of a problem based on the cost and dynamics alone, i.e., without using the policy explicitly. This section follows up on dissipativity theory and shows the existence of a storage functional for the LQR case, hence illustrating the theory presented in this paper. We need to start with the following technical lemma.
\begin{Lemma} \label{Lemma:someMoreShit} Consider the function:
\begin{align}
\label{eq:varsigfunction}
\varsigma\left(\Delta\right) = \Tr{\Delta} - \log\det\left(\Delta + I\right).
\end{align}
For any symmetric matrix $\Delta$ and matrix $M$ such that $\sigma_{\max}\left(M\right) < 1$, the following inequality holds:
\begin{align}
\label{eq:Prestorage}
\left(1-\beta\right)\varsigma\left(\Delta\right) - \varsigma\left(M\Delta M^\top\right) \geq 0,
\end{align}
for any $\beta \leq 1-\sigma_{\max}\left( M\right) $. \flushright$\square$
\end{Lemma}
Equipped with this lemma, we are now ready to provide a storage functional for the LQR case.
\begin{Theorem} \label{LQR:Stability} The choice:
\begin{align}
\lambda\left[\rho_k\right] &=  \kappa\left( \mathrm{Tr}\left(\Sigma_\infty^{-1}\Sigma_k\right)   +\log\mathrm{det}\left(\Sigma_\infty^{-1}\Sigma_k\right) -n \right) \label{eq:LQR:lambda} \nonumber \\
&\quad - \mathrm{Tr}\left(\Omega \left(\Sigma_\infty^{-\frac{1}{2}}\Sigma_k\Sigma_\infty^{-\frac{1}{2}}-I\right)\right)
\end{align}
satisfies the functional dissipativity criterion \eqref{eq:Dissipativity} where matrix $\Omega$ is solution of the discrete Lyapunov equation:
\begin{align}
\label{eq:LyapStorage}
M\Omega M^\top  - \Omega + \Sigma_\infty^{\frac{1}{2}}W\Sigma_\infty^{\frac{1}{2}} = 0,
\end{align}
for $M = \Sigma_{\infty} ^{-\frac{1}{2}}A_\mathrm{c} \Sigma_{\infty} ^{\frac{1}{2}} $ and where $\kappa$, $\varrho$ are constants satisfying:
\begin{align}
\label{eq:DKL:kappa_vartheta}
\kappa  \geq \frac{1}{2\left(1-\sigma_{\max}\left( M\right)\right)},\quad \varrho\leq 2\sigma_{\min}\left(W\right)\sigma_{\min}\left(\Sigma_{\infty}\right).
\end{align}

 \flushright$\square$
\end{Theorem}
\begin{Remark}
It is useful to note here that \eqref{eq:LQR:lambda} cannot be cast as a linear functional of $\rho_k$. This precludes forms like \eqref{eq:LinearRotation}, and confirms the arguments made in the first part of this paper.
\end{Remark}

\subsection{Illustration}
\begin{figure}
\center
\psfrag{Traj}[Bc][Bc][.7]{Trajectories}
\psfrag{Rotatedcost}[Bc][Bc][.7]{Rotated cost}
\psfrag{DKL}[Bc][Bc][.7]{$D_\mathrm{KL}$}
\psfrag{L}[Bc][Bc][.7]{Stage cost}
\psfrag{x}[Bc][Bc][.7]{$\vect s_1$}
\psfrag{y}[Bc][Bc][.7]{$\vect s_2$}
\psfrag{k}[Bc][Bc][.7]{Time step}
\includegraphics[width=.35\textwidth,clip]{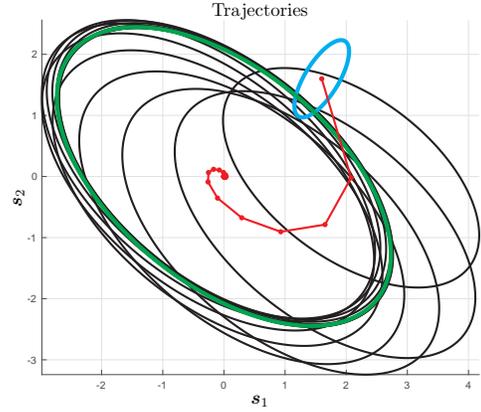}
\caption{Illustration of the LQR case. The expected closed-loop trajectories are depicted in red and the $1 \sigma$ ellipsoids in black. The optimal steady-state density $\rho_\star$ is depicted as the green ellipsoid, and the initial density $\rho_0$ as the black ellipsoid. One can observe how the system converges to $\rho_\star$.
}
\label{fig:Trajectories}
\end{figure}

We illustrate next Theorem \ref{thm:asymptotic_stability}-\ref{LQR:Stability} for the LQR case. We chose a case with $n=2$ states, i.e., $\vect s_k\in\mathbb R^2$, and a scalar action $\vect a_k\in\mathbb R$ having the dynamics
\begin{align}
\vect s_{k+1} = \frac{1}{10}\matr{cc}{\phantom{+}8 & 5 \\ -5 & 7}\vect s_k + \frac{1}{10}\matr{cc}{0 \\ 5}\vect a_k + \vect w_k,
\end{align}
where $\vect w_k \sim \mathcal N\left(0,\Sigma_{\vect w}\right)$ and
\begin{align}
\Sigma_{\vect w} = \matr{cc}{\phantom{+}2 & -1 \\ -1 & 1.6}.
\end{align}
The stage cost is based on the weighting matrices $T = I, R = 1, U = 0$ in \eqref{LQR:cost}. The corresponding constant $L_0$ solution of \eqref{eq:rhoinf} reads as:
\begin{align}
L_0 = 8.92.
\end{align}
The initial density $\rho_0 = \mathcal N\left(\vect \mu_0,\Sigma_0\right)$ was selected, with
\begin{align}
\vect \mu_0 = 1.6\matr{c}{1\\ 1},\quad \Sigma_0 = \frac{1}{10}\matr{cc}{1 & 1 \\ 1 & 2}.
\end{align}
The optimal steady-state density $\rho_\star = \mathcal N\left(\vect 0,\Sigma_\infty\right)$ has the covariance
\begin{align}
\Sigma_\infty = \matr{cc}{\phantom{+}3.73 & -1.76 \\ -1.76& 3}.
\end{align}
The constants $\kappa = 1.72$, $\varrho = 3.14$ then satisfy \eqref{eq:DKL:kappa_vartheta}. Figure~\ref{fig:Trajectories} provides a graphical illustration of this case, showing the trajectories $\rho_k = \mathcal N\left(\vect\mu_k,\Sigma_k\right)$, in terms of their mean (center of the ellipsoids in red) and covariance (ellipsoids). For $k\rightarrow \infty$, the densities converge to the steady-state optimal density $\rho_\star $, depicted as the green ellipsoid here, starting from the initial density $\rho_0$ represented as the light black ellipsoid.

\begin{figure}
\center
\psfrag{Traj}[Bc][Bc][.7]{Trajectories}
\psfrag{Rotatedcost}[Bc][Bc][.7]{Rotated cost $\mathcal L^\mathrm{R}$}
\psfrag{DKL}[Bc][Bc][.7]{$\varphi \cdot D_\mathrm{KL}$}
\psfrag{L}[Bc][Bc][.7]{Stage cost $\mathcal L$}
\psfrag{x}[Bc][Bc][.7]{$\vect s_1$}
\psfrag{y}[Bc][Bc][.7]{$\vect s_2$}
\psfrag{k}[Bc][Bc][.7]{Time step}
\includegraphics[width=.47\textwidth,clip]{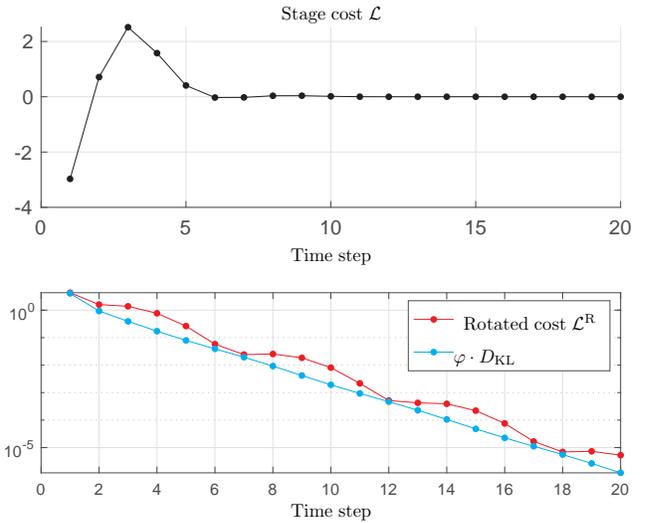}
\caption{Illustration of the LQR case. The upper graph depicts the stage cost \eqref{eq:L:LQR}. The  lower graph depicts $\varphi\cdot D_\mathrm{KL}$ and the rotated cost $\mathcal L^\mathrm{R}$ \eqref{eq:CostModification}, using the storage functional \eqref{eq:LQR:lambda}, using $\varphi=3.13$, selected according to \eqref{eq:DKL:kappa_vartheta}. 
}

\label{fig:DKL}
\end{figure}

Figure \ref{fig:DKL} upper graph shows the stage cost $  \mathcal L\left[\rho_k,\vect\pi\right] $ given by \eqref{eq:L:LQR}. We observe that $\mathcal L$ does not satisfy \eqref{eq:Lbound} as it can take negative values, hence it cannot be used to establish stability as it violates the assumptions of Theorem \ref{thm:asymptotic_stability}. The  lower graph depicts $D_\mathrm{KL}$ (scaled by a factor $\varphi$) and the rotated cost $\mathcal L^\mathrm{R}\left[\rho_k,\vect\pi\right]$ \eqref{eq:CostModification}, using the storage functional \eqref{eq:LQR:lambda} prescribed by Theorem \ref{LQR:Stability}. One can observe how $\mathcal L^\mathrm{R}$ selected as per Theorem \ref{LQR:Stability} is lower-bounded by the scaled $D_\mathrm{KL}$ and satisfies \eqref{eq:Dissipativity}. As a result, it satisfies the conditions of stability \eqref{eq:Lbound} of Theorem \ref{thm:asymptotic_stability}. One can also observe how $D_\mathrm{KL}$ satisfies Proposition \ref{Prop:Generic:Dissimilarity}.

\begin{figure}
\center
\psfrag{Vnorotation}[Bl][Bl][.7]{$J_\star$ from \eqref{eq:MDP}}
\psfrag{Vrotationofcost}[Bl][Bl][.7]{$V^R_\star$ from \eqref{eq:rotatetd_FOCP}}
\psfrag{k}[Bc][Bc][.7]{Time step}
\psfrag{Vall}[Bc][Bc][.7]{$V_\star^R$ and $J_\star$}
\includegraphics[width=.47\textwidth,clip]{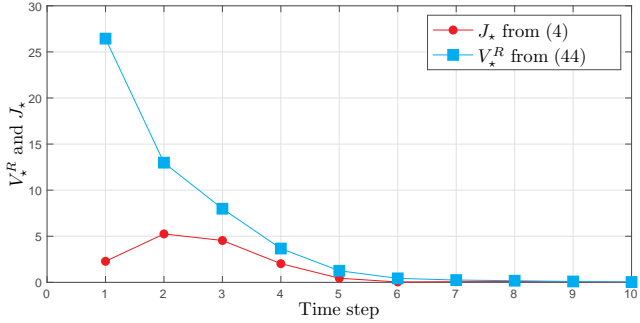}
\caption{Illustration of the LQR case. The red curve represents the evolution of the value function $J_\star$ from the original problem  \eqref{eq:MDP}. 
The black curve represents the evolution of the value function $V_\star^\mathrm{R}$ of the rotated problem \eqref{eq:rotatetd_FOCP}, which 
is a Lyapunov function for the system, as established in Theorem~\ref{Th:Stability}.
}
\label{fig:V}
\end{figure}

Figure \ref{fig:V} illustrates the evolution of the value function $J_\star$ from the original MDP  \eqref{eq:MDP} (see red curve). One can readily see that $J_\star$ is not a Lyapunov function as it is not decreasing. The cyan curve represents the evolution of the value function $V_\star^\mathrm{R}$ of the rotated problem \eqref{eq:rotatetd_FOCP}, which decreases monotonically, as {established} by Theorem \ref{Th:Stability}.

\begin{figure}
\center
\psfrag{DKL}[Bl][Bl][.7]{$22.7\cdot D_{\mathrm{KL}}$}
\psfrag{Vrotationofcost}[Bl][Bl][.7]{$V^R_\star$ from \eqref{eq:rotatetd_FOCP}}
\psfrag{k}[Bc][Bc][.7]{Time step}
\psfrag{Vall}[Bc][Bc][.7]{$V_\star^R$ and $D_{\mathrm{KL}}$}
\includegraphics[width=.47\textwidth,clip]{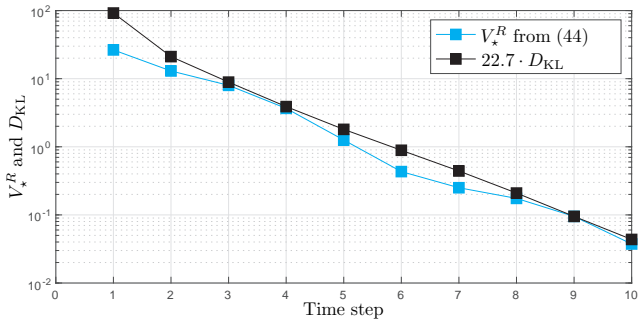}
\caption{Illustration of the LQR case. The black curve represents the evolution of the value function $V_\star^\mathrm{R}$ of the rotated problem \eqref{eq:rotatetd_FOCP}. The black curve is $D_{\mathrm{KL}}$ scaled by a factor $22.7$, and is upper-bounding $V_\star^\mathrm{R}$, illustrating assumption \eqref{eq:controllability} in Theorem~\ref{Th:Stability} on these specific trajectories and specific case.
}
\label{fig:VDKL}
\end{figure}

Figure \ref{fig:VDKL} illustrates assumption \eqref{eq:controllability} in Theorem~\ref{Th:Stability}, and similarly assumption \eqref{eq:Vbound} in Theorem \ref{thm:asymptotic_stability}. As detailed above, these assumptions are difficult to verify formally even in the deterministic case. One can observe, however, that they hold on this specific example and on the proposed trajectories.


\section{Conclusion}
\label{sec:conclusions}
In this paper, we proposed a generalization of the established economic MPC dissipativity theory to Markov Decision Processes. We explain why this generalization is not straightforward, and show that it can be done by extending the notion of storage functions to nonlinear storage functionals. A classic Lyapunov argument can then be used to discuss the asymptotic stability of the probability measures underlying the Markov Decision Processes to the steady-state optimal probability measure. The asymptotic stability can be expressed in terms of dissimilarity measures such as the Kullback-Leibler divergence, the Wasserstein metric, or the total variational distance. The theory is illustrated on the LQR case with Normal process noise, for which a storage functional can be explicitly provided. 

Current work extends this theory to discussing the stability of Stochastic MPC, the construction of stability-constrained learning using MPC, and the extension to the discounted case, by building on the theory proposed in~\cite{Zanon2021d}. Furthermore, an extension to policies based on finite-horizon MPC schemes will be considered.

\bibliographystyle{plain}
\bibliography{bibliography}

\section{Appendix}
The proofs of Lemma \ref{Lemma:Bazdmeg}, Lemma \ref{Lemma:ConvergenceInEigenvalues}, Proposition \ref{Prop:Generic:Dissimilarity}, Lemma \ref{Lemma:someMoreShit}, and Theorem \ref{LQR:Stability} 
are provided hereafter. Note that we will denote the ordered eigenvalues of matrix $A$ as $\Lambda_{1,\ldots,n}(A)$ and its ordered singular values as $\sigma_{1,\ldots,n}(A)$. We will denote the trace of $A$ as $\mathrm{Tr}\left(A\right)$. We will further use:
\begin{align*}
	\Lambda_i(AB) &= \Lambda_i(BA),& 1 + c\Lambda_i(A)	 &= \Lambda_i(I + cA), \\
	\Lambda_i(AA^\top) &= \Lambda_i(A)^2, & \max_k \sigma_k(A)\sigma_i(B) &\geq \sigma_i(AB) , \\
	\mathrm{Tr}(ABC)&=\mathrm{Tr}(CAB), & \det(A)\det(B)&=\det(AB), \\
	\det(ABC)&=\det(CAB)
\end{align*}
and all such permutations
\subsection{Proof of Lemma \ref{Lemma:Bazdmeg}}
First, we observe that the following inequality holds:
\begin{align}
\sigma\left(M\Delta M^\top\right) \leq \sigma_{\max}\left(M\right)^2\sigma\left(\Delta\right), 
\end{align}
where $\sigma\left(\cdot\right)$ is the vector of singular values. Hence since $\Delta$ is symmetric: 
\begin{align}
\Lambda_i\left(M\Delta M^\top\right)^2 \leq \sigma_{\max}\left(M\right)^2  \Lambda_i\left(\Delta\right)^2,
\end{align}
such that 
\begin{align}
\left|\Lambda_i\left(M\Delta M^\top\right)\right| = |\alpha_i| \left|  \Lambda_i\left(\Delta\right)\right|
\end{align} 
holds for some sequence $|\alpha_{1}|,\ldots,|\alpha_n| \leq \sigma_{\max}\left(M\right)$. We then need to show that  $\alpha_{i}>0$, $i = 1,\ldots,n$ . We observe that:
\begin{align}
\Lambda_i\left(M\Delta M^\top\right)  = \Lambda_i\left(\Phi \Delta\right),
\end{align}
where $\Phi =  M^\top M$ is symmetric, positive definite. Let us define:
\begin{align}
\Gamma(t) = e^{t\log \Phi}\Delta,
\end{align}
where we use the matrix exponential and logarithms. Then
\begin{align}
\Gamma(0) = \Delta\quad\text{and}\quad \Gamma(1) =  M^\top M \Delta
\end{align}
trivially hold. We further observe that
\begin{align}
\Det{\Gamma(t)}  &= \Det{e^{t \log \Phi}}\Det{\Delta} = e^{t\Tr{ \log \Phi}}\Det{\Delta}  \\
&= \left(e^{\Tr{ \log \Phi}}\right)^t\Det{\Delta}= \Det{ \log \Phi}^t\Det{\Delta}. \nonumber
\end{align}
Since $\Det{\Gamma(1)} = \Det{ M \Delta M^\top} \neq 0$ by assumption, it follows that $\Det{\Gamma(t)} \neq 0$ for all $t\in[0,1]$. We can then conclude that the eigenvalues
\begin{align}
\Lambda_i\left(\Gamma(t)\right)  \neq 0,\quad \forall\, t,
\end{align}
such that $\Lambda_i\left(\Gamma(t)\right)$ does change sign over $t\in[0,1]$. This establishes that  $\alpha_{i}>0$, $i = 1,\ldots,n$  and hence \eqref{eq:BazdmegInequalities}.
\subsection{Proof of Lemma \ref{Lemma:ConvergenceInEigenvalues}}
We first observe that the dynamics for $\Sigma_k$ can be reformulated as:
\begin{align}
	\label{eq:Psi:Dynamics}
\Psi_{k+1} = M\Psi_k M^\top + N,
\end{align}
where $\Psi_k = \Sigma_{\infty} ^{-\frac{1}{2}}\Sigma_{k}\Sigma_{\infty} ^{-\frac{1}{2}} $ and
\begin{align}
\label{eq:MandN:def}
M = \Sigma_{\infty} ^{-\frac{1}{2}}A_\mathrm{c} \Sigma_{\infty} ^{\frac{1}{2}} ,\qquad N = \Sigma_{\infty} ^{-\frac{1}{2}}\Sigma_{\vect w}\Sigma_{\infty} ^{-\frac{1}{2}},
\end{align}
such that $\lim_{k\rightarrow \infty } \Psi_k = I$ and
\begin{align}
MM^\top + N &= I.
\end{align}
We can then observe that
\begin{align}
\sigma_i\left(M\right)^2 &= \Lambda_i\left(M^\top M\right) = \Lambda_i\left(MM^\top\right) = \Lambda_i\left(I-N\right) \nonumber \\& = 1-\Lambda_i(N) \geq 0,
\end{align}
and that
\begin{align}
\Lambda_i\left(N\right) &= \Lambda_i\left(\Sigma_{\infty} ^{-\frac{1}{2}}\Sigma_{\vect w}\Sigma_{\infty} ^{-\frac{1}{2}}\right) = \Lambda_i\left(\Sigma_{\infty} ^{-1}\Sigma_{\vect w}\right) > 0,
\end{align}
since $\Sigma_{\infty}$, $\Sigma_{\vect w}$ are positive definite. It follows that 
\begin{align}
\sigma_i\left(M\right) \in [0,1).
\end{align}
  Let us label
  \begin{align}
\Delta_k = \Psi_k-I,
\end{align}
and observe that
\begin{align}
	\label{eq:Delta:Dynamics}
\Delta_{k+1} = M\Delta_k M^\top \quad\text{and}\quad \Lambda_i\left(\Delta_k\right) = \Lambda\left(\Psi_k\right) - 1.
\end{align}
Using Lemma \ref{Lemma:Bazdmeg}, we observe that \eqref{eq:BazdmegInequalities} applies, i.e.
\begin{align}
\Lambda_i\left( \Delta_{k+1} \right) = \alpha_i\Lambda_i\left(\Delta_k \right),\quad i=1,\ldots,n,
\end{align} 
for a sequence  $\alpha_{i}>0$, $i = 1,\ldots,n$  with $\alpha_i \leq \sigma_{\max}\left(M\right)< 1$. Finally, we observe that:
\begin{align}
\Lambda_i\left(\Delta_k\right) = \Lambda_i\left(\Sigma_{\infty} ^{-\frac{1}{2}}\Sigma_{k}\Sigma_{\infty} ^{-\frac{1}{2}} -I\right) = \Lambda_i\left(\Sigma_{\infty}^{-1} \Sigma_{k}\right) -1,
\end{align}
and conclude that \eqref{eq:Contraction:Eigen} holds.

\subsection{Proof of Proposition \ref{Prop:Generic:Dissimilarity}}
We first observe that the monotonic convergence of the second term in \eqref{eq:Generic:Dissimilarity}
\begin{align}
\sum_{i=1}^n \zeta\left(\Lambda_i\left(\Sigma_\infty^{-1}\Sigma_k\right)\right)
\end{align}
follows directly from Lemma \ref{Lemma:ConvergenceInEigenvalues}. The convergence of the first term follows classic system dynamic theory. We recall the argument for completeness. 
Consider the state space transformation:
\begin{align}
\vect\nu_k &= \Sigma_{\infty} ^{-\frac{1}{2}} \vect \mu_k, 
\end{align}
following the dynamics:
\begin{align}
\vect\nu_{k+1} &= M\vect\nu_k,
\end{align}
with 
\begin{align}
\Lambda_i\left(M^\top M\right)  \leq  \sigma_{\max}\left(M \right)^2 < 1.
\end{align}
It follows that
\begin{align}
\vect\mu_{k+1}^\top \Sigma_\infty^{-1}\vect\mu_{k+1}&=\|\vect\nu_{k+1}\|^2 =  \vect\nu_k^\top M^\top M\vect\nu_k < \|\vect\nu_k\|^2 \nonumber \\&= \vect\mu_{k}^\top \Sigma_\infty^{-1}\vect\mu_{k}.
\end{align}
\subsection{Proof of Lemma \ref{Lemma:someMoreShit}}
We first observe that for any $a < 1$ the inequality:
\begin{align}
\label{eq:SillyIneq}
&\vartheta_{a,b}(x) \\
&\hspace{0.5em}:= \left(1-b\right)\left(x - \log \left(x+1\right)\right) - ax + \log \left(ax+1\right)  \geq 0\nonumber
\end{align}
holds on $x\in(-1,\infty)$ for $0 < b \leq 1-a < 1$. Indeed, we observe that $\vartheta_{a,b}(0)=0$ and that on the interval $x\in (-1,\infty)$
\begin{align}
\frac{\mathrm{d} \vartheta_{a,b}}{\mathrm d x} = -\frac{ax\left(\left(a + b -1\right)x - 1 + b\right)}{\left(ax+1\right)\left(x+1\right)} = 0
\end{align}
has the unique solution $x=0$. Furthermore, the sign of $\frac{\mathrm{d} \vartheta_{a,b}}{\mathrm d x}$ entails that $\vartheta_{a,b}$ is monotonically increasing away from $x=0$, which establishes \eqref{eq:SillyIneq}. Using Lemma \ref{Lemma:Bazdmeg}, we then observe that for all $i$:
\begin{align}
\Lambda_i\left( M\Delta M^\top \right) = \Lambda_i\left(M^\top M\Delta \right) = \alpha_i\Lambda_i\left(\Delta \right)
\end{align}
holds for some sequence $\alpha_{1,\ldots,n} \leq  \sigma_{\max}\left( M\right)$. Then 
\begin{align}
& - \Lambda_i\left( M\Delta M^\top\right) + \log\left(\Lambda_i\left( M\Delta M^\top \right)+1\right) = \\
& - \alpha_i\Lambda_i\left( \Delta \right) + \log\left(\alpha_i\Lambda_i\left( \Delta  \right)+1\right). \nonumber
\end{align}
Hence, using 
\begin{align}
\varsigma\left(\Delta\right) &= \sum_{i=1}^n \Lambda_i\left(\Delta\right) - \log\left(\Lambda_i\left(\Delta\right)+1\right)  \\
\varsigma\left(M\Delta M^\top\right) &= \sum_{i=1}^n \Lambda_i\left( M\Delta M^\top\right) \nonumber\\
&\hspace{4em}- \log\left(\Lambda_i\left( M\Delta M^\top \right)+1\right), \nonumber
\end{align}
we observe that
\begin{align}
&\left(1-\beta\right)\varsigma\left(\Delta\right) - \varsigma\left(M\Delta M^\top\right) =
\sum_{i=1}^n\vartheta_{\alpha_i,\beta}(\Lambda_i\left( \Delta  \right)), \nonumber
\end{align}
such that the choice
\begin{align}
\beta \leq \min_i 1-\alpha_i \leq 1-\sigma_{\max}\left( M\right) < 1
\end{align}
ensures that \eqref{eq:Prestorage} holds.

 \subsection{Proof of Theorem \ref{LQR:Stability}}
We first observe that $\lambda\left[\rho_\star\right] = 0$ holds by construction. We further observe that:
\begin{align}
\mathrm{Tr}\left(\Omega \left(\Sigma_\infty^{-\frac{1}{2}}\Sigma_k\Sigma_\infty^{-\frac{1}{2}}-I\right)\right) = \mathrm{Tr}\left(\Omega \Delta_k\right),
\end{align}
and, using \eqref{eq:Delta:Dynamics} and \eqref{eq:LyapStorage}, we obtain:
\begin{align}
\label{eq:KillingTerm}
\mathrm{Tr}\left(\Omega \Delta_k\right) -\mathrm{Tr}\left(\Omega \Delta_{k+1}\right) 
&= -\mathrm{Tr}\left( W \left ( \Sigma_k-\Sigma_\infty \right ) \right).
\end{align}
Using \eqref{eq:KillingTerm} and \eqref{eq:varsigfunction} we then observe that:
\begin{align}
\label{eq:SumOfLambda}
\lambda\left[\rho_{k}\right] - \lambda\left[\rho_{k+1}\right] &= \kappa(\varsigma\left(\Delta_k\right)-\varsigma\left(\Delta_{k+1}\right)) \\ &\qquad - \mathrm{Tr}\left( W \left ( \Sigma_k-\Sigma_\infty \right ) \right). \nonumber
\end{align}
Using \eqref{eq:L:LQR} and \eqref{eq:SumOfLambda} and Lemma \ref{Lemma:someMoreShit}, it follows that
\begin{align}
&\mathcal L\left[\rho_k,\vect\pi[\rho_k]\right] - \lambda\left[\rho_{k+1}\right] + \lambda\left[\rho_k\right]  \\ 
&\hspace{1cm} =\vect\mu_k^\top W\vect\mu_k +  \kappa(\varsigma\left(\Delta_k\right)-\varsigma\left(\Delta_{k+1}\right))\nonumber \\
&\hspace{1cm}\geq \vect\mu_k^\top W\vect\mu_k +  \kappa \beta \varsigma\left(\Delta_k\right). \nonumber
\end{align}
We further observe that
\begin{align}
 D_{\mathrm{KL}}\left(\rho_k\, ||\, \rho_\star\right) =&\frac{1}{2}\vect\mu_k^\top \Sigma_{\infty}^{-1}\vect\mu_k  + \frac{1}{2}\varsigma\left(\Delta_k\right).
\end{align} 
It follows that for \eqref{eq:DKL:kappa_vartheta}, $\kappa \geq\frac{1}{2\beta}$, and the inequality
\begin{align}
\mathcal L\left[\rho_k,\vect\pi[\rho_k]\right] - \lambda\left[\rho_{k+1}\right] + \lambda\left[\rho_k\right] \geq \varrho  D_{\mathrm{KL}}\left(\rho_k\, ||\, \rho_\star\right)
\end{align}
holds.

\end{document}